\documentclass[12pt]{article}
\usepackage{times}
\usepackage{geometry}
\usepackage{graphicx}
\usepackage[utf8]{inputenc}
\usepackage{enumitem,amssymb}
\usepackage{ragged2e}
\usepackage{wrapfig}
\usepackage{color}
\newlist{thematic}{itemize}{8}
\setlist[thematic]{label=$\square$}
\usepackage{pifont}

\newcommand{\arcsec}{^{\prime\prime}}

\begin{document}
{\raggedright
\huge
Astro2020 Science White Paper \linebreak

Cold Debris Disks as Strategic Targets for the 2020s} \linebreak
\normalsize

\noindent \textbf{Thematic Areas:} \hspace*{60pt} $\boxtimes$ Planetary Systems \hspace*{10pt} $\square$ Star and Planet Formation \hspace*{20pt}\linebreak
$\square$ Formation and Evolution of Compact Objects \hspace*{31pt} $\square$ Cosmology and Fundamental Physics \linebreak
  $\square$  Stars and Stellar Evolution \hspace*{1pt} $\square$ Resolved Stellar Populations and their Environments \hspace*{40pt} \linebreak
  $\square$    Galaxy Evolution   \hspace*{45pt} $\square$             Multi-Messenger Astronomy and Astrophysics \hspace*{65pt} \linebreak
  
\noindent\textbf{Principal Author:}

\noindent Name: John H. Debes \\	
Institution:  Space Telescope Science Institute \\
Email: debes@stsci.edu \\
Phone:  (410)338-4782 \\

\noindent\textbf{Co-authors:} 
Elodie Choquet (Aix Marseille Univ, CNRS, CNES, LAM),Virginie C. Faramaz (JPL-Caltech), Gaspard Duchene (Berkeley), Dean Hines (STScI), Chris Stark (STScI), Marie Ygouf (Caltech-IPAC), Julien Girard (STScI), Amaya Moro-Martin (STScI),  Pauline Arriaga (UCLA), Christine Chen (STScI), Thayne Currie (NASA/Ames, NAOJ), Sally Dodson-Robinson (U. Delaware), Ewan S. Douglas (MIT/U. of Arizona), Paul Kalas (Berkeley), Carey M. Lisse (APL), Dimitri Mawet (Caltech), Johan Mazoyer (JPL-Caltech), Bertrand Mennesson (JPL), Max A. Millar-Blanchaer (JPL-Caltech), Anand Sivramakrishnan (STScI), Jason Wang (Caltech) \\

\noindent\textbf{Co-Signers:} Vanessa Bailey (JPL), William C. Danchi (GSFC), Laurent Pueyo (STScI), Marshall Perrin (STScI), Bin Ren (STScI), Aki Roberge (GSFC), Glenn Schneider (U. of Arizona), Jordan Steckloff (PSI) \\

\noindent\textbf{Abstract:}
Cold debris disks (T$<$200~K) are analogues to the dust in the Solar System's Kuiper belt--dust generated from the evaporation and collision of minor bodies perturbed by planets, our Sun, and the local interstellar medium. Scattered light from debris disks acts as both a signpost for unseen planets as well as a source of contamination for directly imaging terrestrial planets, but many details of these disks are poorly understood. We lay out a critical observational path for the study of nearby debris disks that focuses on defining an empirical relationship between scattered light and thermal emission from a disk, probing the dynamics and properties of debris disks, and directly determining the influence of planets on disks. 

We endorse the findings and recommendations published in the National
Academy reports on Exoplanet Science Strategy and Astrobiology Strategy for
the Search for Life in the Universe. This white paper extends and complements
the material presented therein with a focus on debris disks around nearby stars. Separate complementary papers are being submitted regarding the inner warm regions of debris disks (Mennesson et al.), the modeling of debris disk evolution (Gaspar et al.), studies of dust properties (Chen et al.), and thermal emission from disks (Su et al.).

\pagebreak

\subsection*{Overview}


As we approach the detection of the 4000th confirmed exoplanet, we are on the verge of foundational discoveries in planet formation, habitability and the context of our Solar System within the galactic planetary population (Exoplanet Science Strategy report 2018; Plavchan et al. 2019). Direct imaging of planetary systems provides key insights on their architecture, chemistry, dynamics and evolution. 


Debris disks are the dusty remains of planet formation--they trace the location of colliding or evaporating minor bodies in a planetary system. They can be detected in scattered visible/near-infrared (NIR) light and via thermal emission in the IR as well as sub-mm wavelengths. Thermal emission directly traces larger particles in the system and determines the temperature of the dust, while the scattered light traces the micron-sized dust. We define a cold debris disk as a disk with a blackbody temperature $\leq$200~K. 

A proxy for the total amount of dust in a system can be estimated from the ratio of thermal infrared radiation to the total luminosity of the star or $L_{\rm IR}/L_{\star}$. The dust properties and disk morphology dictate the peak surface brightness of scattered light. The dynamics of micron-sized dust are highly dependent on what non-gravitational forces are present, while larger dust particles trace the density distribution of parent bodies. Further, more dense disks are dominated by collisions (so-called collisional disks), while less dense disks are shaped by drag forces, such as Poynting-Robertson drag (so-called transport dominated disks). To date most disks that have been spatially resolved in scattered light are cold and collision dominated. The interplay between dynamical forces and dust particle properties makes predicting the morphology and peak surface brightness for dusty disks challenging. This uncertainty hinders the detection of unseen planets within disks and planning for future exo-Earth yields from direct imaging missions.

\begin{figure}[t]
\begin{minipage}[b]{0.46\linewidth}
\centering
\includegraphics[width=\textwidth]{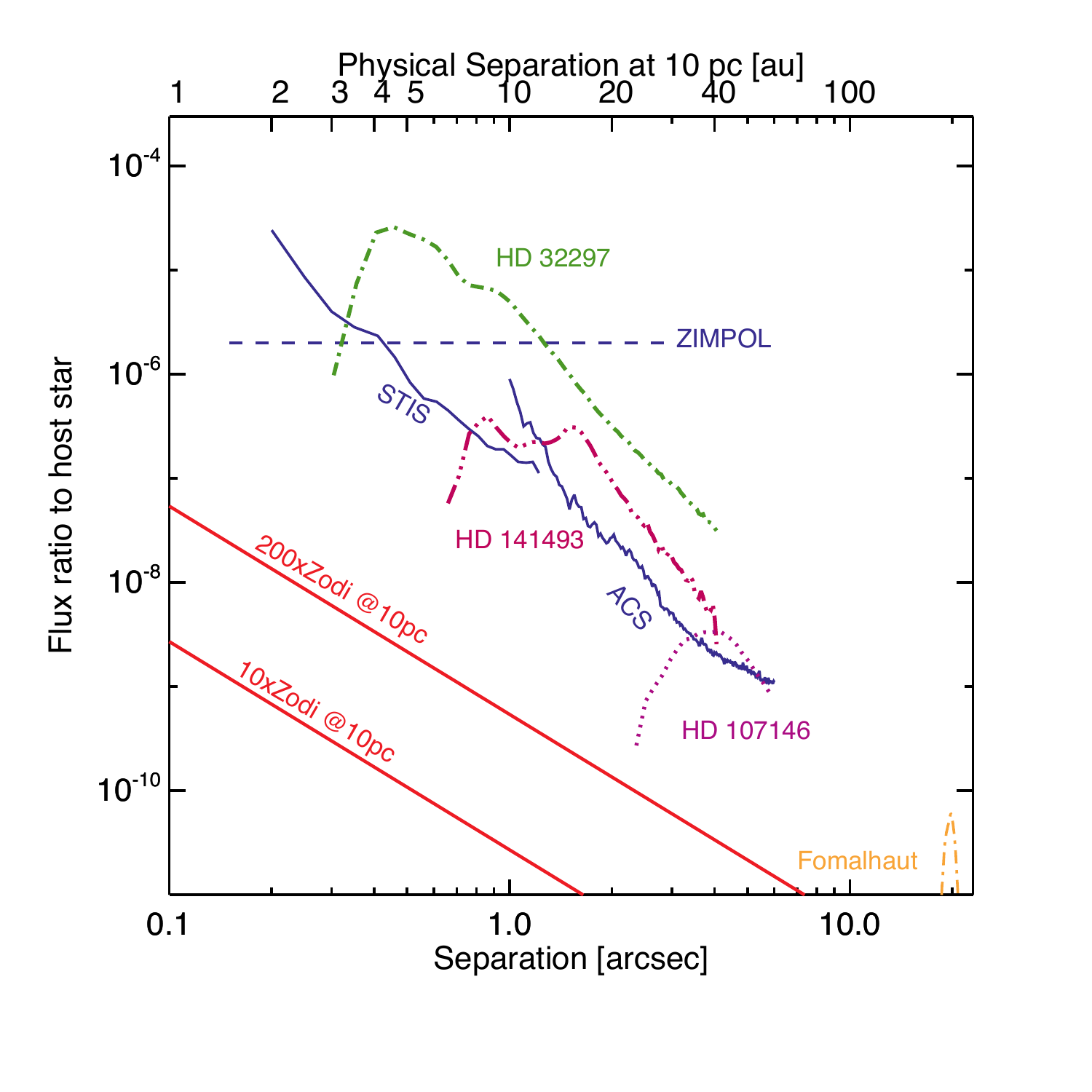}
\end{minipage}
\hspace{0.1cm}
\begin{minipage}[b]{0.46\linewidth}
\centering
\includegraphics[width=\textwidth]{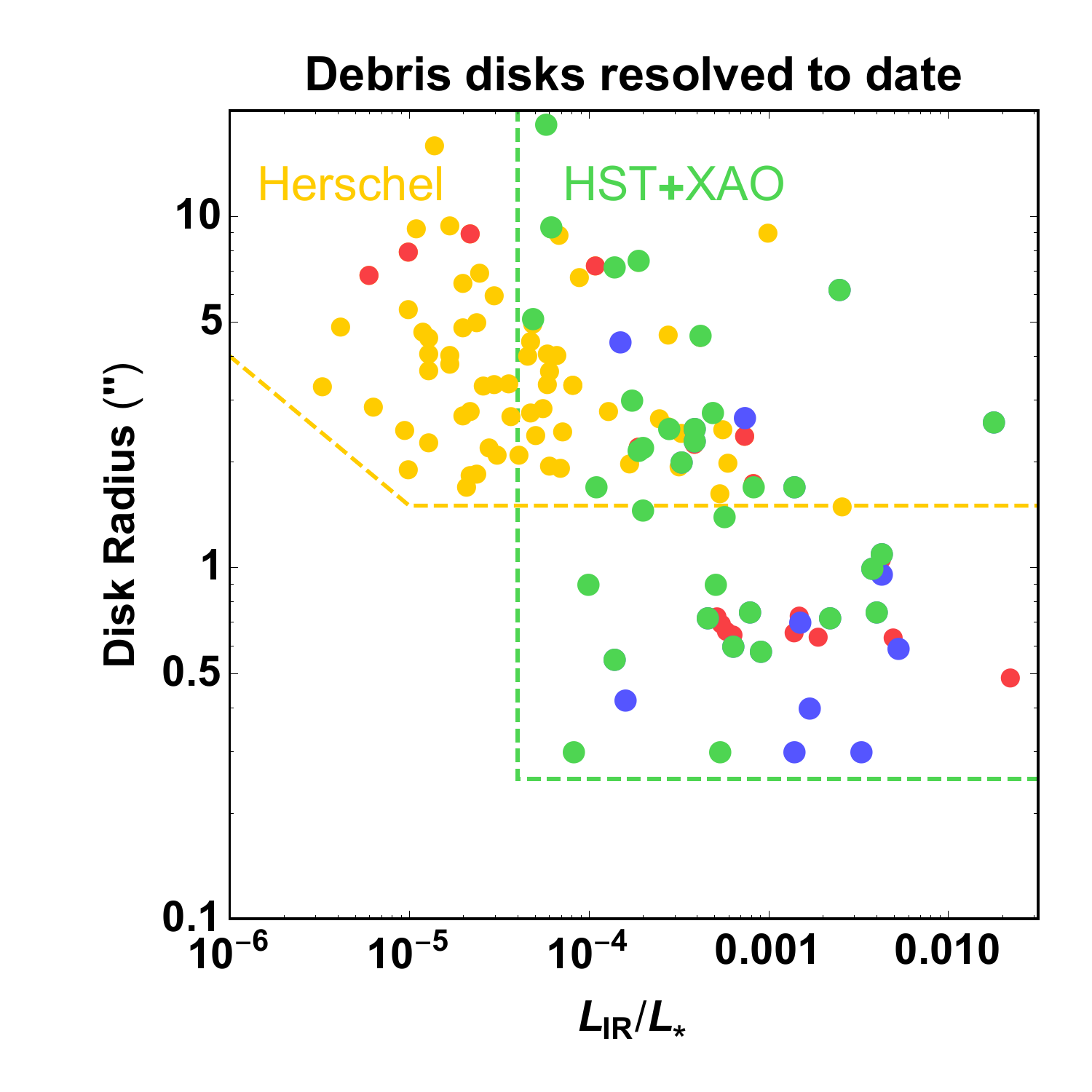}
\end{minipage}
\caption{(Left) Surface brightness profiles along the disk major axis of a selection of spatially resolved disks in scattered light, along with detection limits from visible light coronagraphs. (Right) Radii and $L_{\rm IR}/L_{\star}$ of the cold debris disks resolved to date with different facilities including the Hubble Space Telescope (HST) instruments and extreme-AO (XAO). The brightest systems have been resolved in scattered light (green and blue markers) while a majority of {\em Herschel}-resolved disks (yellow markers) lie below the sensitivity limits of current visible/near-IR imagers.}
\label{fig:f1}
\end{figure}

Scattered light images of cold debris disks are useful because they tend to have better spatial resolution compared to thermal IR or sub-mm interferometric imaging as well as tracing the warmest and coldest components of a disk with a single wavelength of light. Cold debris disks are more extended and can thus be probed to a larger volume, improving the chances of building a statistically significant sample with a wide range of morphology, host star, and $L_{\rm IR}/L_{\star}$.

We recommend that three areas of cold debris disk research be prioritized over the next decade in order to connect existing observations to planned direct imaging missions: predicting scattered light emission from disks based on their thermal emission, characterizing the dynamics and properties of dust in debris disks, and forging a connection between exoplanets and structures observed in debris disks. Focusing on these areas will maximize the main scientific goals of the Exoplanet Science Strategy. Throughout this report we mention selected current and planned facilities that can accomplish a subset of these goals, but we primarily focus on scientific requirements rather than specific facilities.

\subsection*{Defining a relationship between scattered light emission and thermal emission}

Cold debris disks are relatively common in the solar neighborhood. According to IR excesses surveys performed with {\em Herschel}, debris disks with $L_{\rm IR}/L_{\star}\gtrsim 10^{-6}$ are as common as 22\% around K--A stars (Montesinos et al. 2016) and are detected around stars as old as 10 Gyr. About a hundred debris disks have been resolved in thermal emission with Herschel, JCMT, or ALMA on spatial scales as small as $0.3''$ (e.g. Lieman-Sifry. 2016), overlapping with the regions where we expect to caracterize exoplanets in reflected light. The current sample of spatially resolved scattered light debris disks numbers to $\sim40$, the majority discovered via instruments on the Hubble Space Telescope (HST) (Choquet et al. 2018) and with $L_{\rm IR}/L_{\star}\sim10^{-4}$. Disks with lower luminosities are undetectable with current facilities, but the exact amount of scattered light they may possess is difficult to predict.

Disk luminosities below $L_{\rm IR}/L_{\star}\sim10^{-4}$ correspond roughly to the transition point between collision dominated disks and transport dominated disks, weakly dependent on disk radius (Kuchner \& Stark, 2010). A majority of nearby disks possess a fractional luminosity that suggests that they may be transport dominated (Chen et al. 2014, Sibthorpe et al. 2018). Transport dominated disks are of particular interest because they should be cold analogs to the Kuiper belt and Solar Sytem zodiacal cloud. Further, it is important to empirically determine how scattered light efficiency evolves with decreasing $L_{\rm IR}/L_{\star}$. Determining this relationship will help estimate dust scattered light contamination at the brightness levels and spatial scales where we expect to characterize the reflected light of exoplanets.


The superior inner working angles of ground-based extreme AO coronagraphs will find additional bright disks (e.g., Engler et al., 2018) without improving on HST's detection limits at larger separations (e.g. Marshall et al. 2018).
Figure \ref{fig:f1} shows the surface brightness profiles of disks as a function of angular distance from their host stars. We also plot extrapolated surface brightness profiles of hypothetical disks that have surface brightnesses 10 and 200 times as bright as that estimated for the Solar System exo-zodiacal cloud at 1~AU at a distance of 10~pc as proxies for disks that may be transport dominated. These disks demonstrate the diversity of surface brightnesses with respect to $L_{\rm IR}/L_{\star}$ along with the challenges these disks present for the direct imaging of exoplanets in reflected light.

We recommend a systematic and homogenous scattered light direct imaging survey of {\em Herschel} detected disks with low luminosities to empirically determine the amount of scattered light as it relates to integrated thermal emission. This survey will require both high sensitivity at small inner working angles to resolve the most compact disks, but will also require a surface brightness sensitivity of at least V=25 mag arcsec$^{-2}$ at 5-10$\arcsec$. We also recommend that concurrent to this effort that each star have extremely high precision spectro-photometry in the visible/NIR performed. Such observations allow the search for integrated scattered light beyond 2$\mu$m as well as accurate estimation of stellar photospheres. While IRTF/SPEX can detect integrated scattered light from the brightest disks (Lisse et al. 2012; 2017), a relative flux precision of better than at least one part in 10$^{5}$ will be needed for most {\em Herschel} detected disks. For larger disks, JWST's NIRCAM will likely be preferred for cataloging scattered light beyond XAO NIR capabilities.




\subsection*{Probing the dynamics and properties of dust with scattered light}



Dust interacts with other dust, with its host star, and with its local galactic environment in ways that are observable. These include variations in morphology between scattered light and thermal emission, differences in scattering efficiency and polarization, and time variable sub-structures (Boccaletti et al., 2016). Interpreting the origins of these observable effects will be critical to predicting the amount of scattered light from a disk as well as when it is being modified by a planet.

Observations of debris disks in scattered-light and at millimeter wavelengths are crucial for understanding what dynamical forces act upon dust. Historically, observations of asymmetries in debris disks seen in scattered light have been attributed to planetary perturbers, but the discovery of several disks that show surprising morphologies in scattered light and in sub-mm wavelengths challenge these simple interpretations.


As the sample of debris disks observed both in scattered light and at millimeter wavelengths increases (e.g., Boley et al. 2012) some disks are equally broad in scattered light and thermal emission, contrary to expectations. For instance, the millimeter emission of the debris ring of HD 107146 (Ricci et al. 2015, Marino et al. 2018) and of that of HD 92945 (Marino et al. 2019) seen with ALMA both appear as broad as seen in scattered-light. It is difficult to explain differences on the disks morphologies and width across wavelengths due solely to radiation pressure, which means that our understanding of the dust dynamics is still incomplete. Our proposed survey of Herschel-detected disks will constrain the extent to which stellar wind drag, ISM interactions, and other dynamical effects may affect the morphology of a debris disk.

 Scattered light images of debris disks also provide clues about the physical characteristics of the dust grains they contain, which in turn inform on the nature of the parent bodies from which the grains are produced. A complementary white paper by Chen et al. also discusses the importance of scattered light imaging for understanding dust composition. Quantitatively, the grain size distribution, dust composition (including the possibility of porosity) and the shape of individual grains (oblate/prolate grains, aggregates) is encoded in the scattered light spectrum, scattering phase function (total intensity light as a function of scattering angle) and polarizability curves (linear polarized intensity of light as a function of scattering angle). The study of scattering efficiency as a function of wavelength has been studied for over a decade in debris disks (e.g. Debes et al., 2008) and the field is now moving toward estimating all three quantities simultaneously. 

Cold debris disks and a wide range of Solar System bodies show modest differences between phase functions (Hughes et al. 2018). Indeed, in the Solar System, studies of cometary dust primarily focus on the polarizability curve, whose main characteristics (peak level of polarization, amplitude of the backscattering negative branch) appear more dependent on dust properties (e.g., Levasseur-Rigourd 2001; Zubko et al. 2016; Kolokolova et al. 2004, 2018; Frattin et al. 2019). For instance, the lack of a negative branch altogether in near-infrared observations of comet Hale-Bopp (Jones \& Gehrz 2000) was linked to the remarkably small size of the dust grains produced in that system. 

\begin{figure}
\begin{minipage}[b]{0.45\linewidth}
\centering
\includegraphics[width=\textwidth]{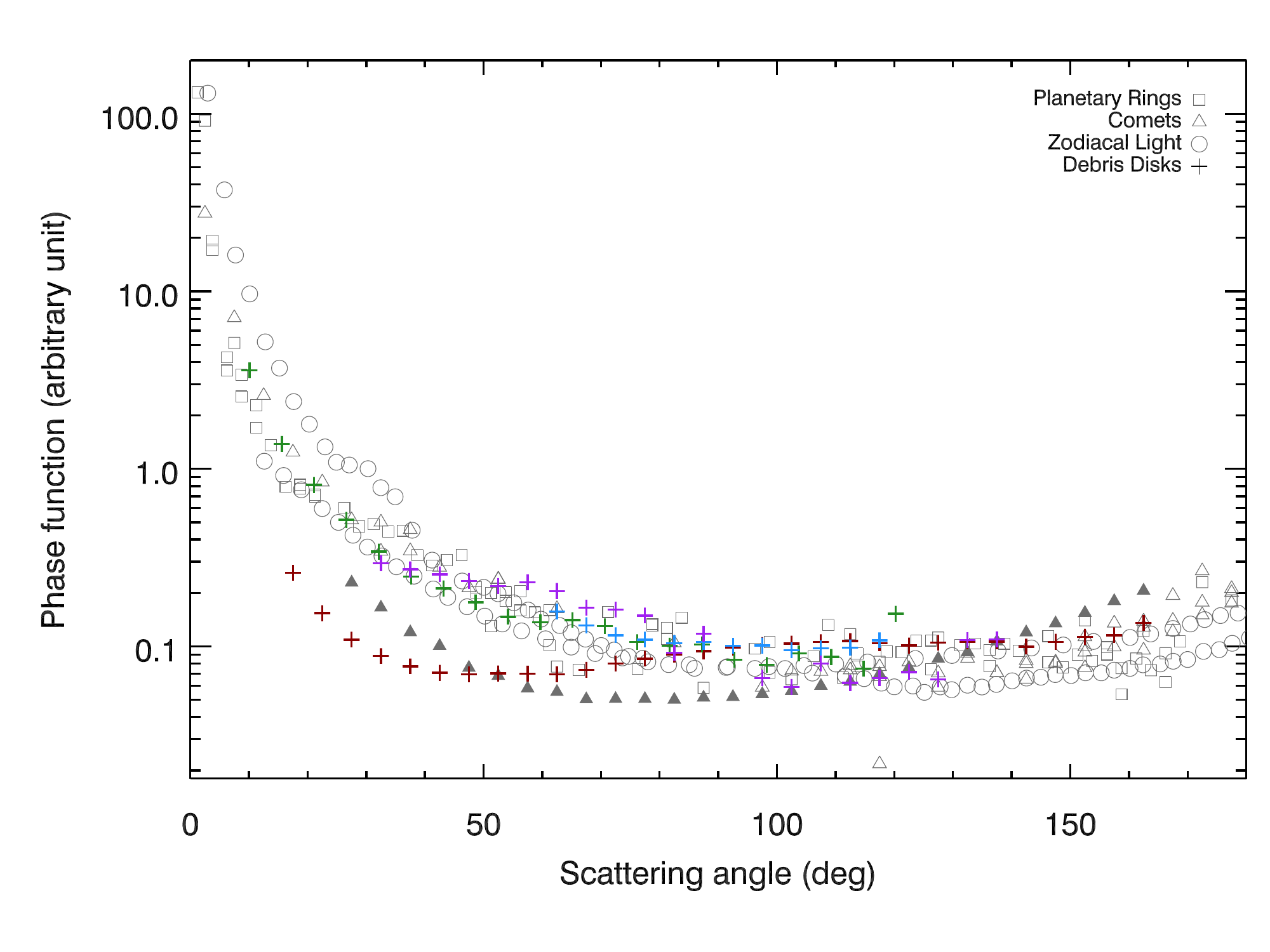}
\end{minipage}
\hspace{0.5cm}
\begin{minipage}[b]{0.45\linewidth}
\centering
\includegraphics[width=\textwidth]{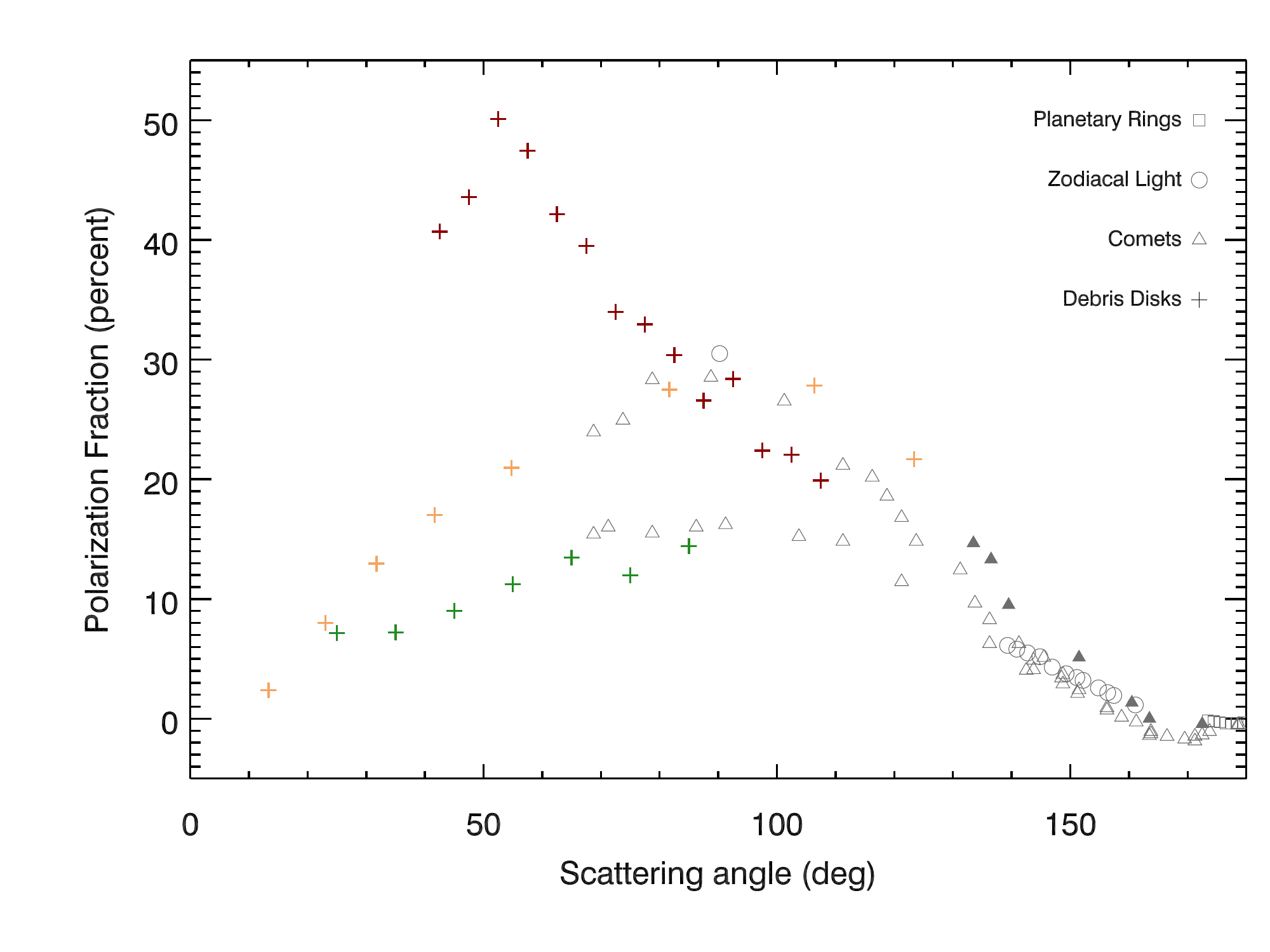}
\end{minipage}
\caption{Total intensity (left, adapted from Hughes et al. 2018) and polarizability (right) for Solar System dust populations (gray symbols) and for debris disks (colored symbols). Most phase functions share a near universal phase function, except for the HR\,4796\,A debris and comet 67P, while the scatter in polarizability curves is much larger (filled triangles represent comet Hale-Bopp for which the negative polarization backscattering branch is conspiscuously absent.)}
\label{fig:dustscatt}
\end{figure}

We recommend that a large sample of disks be observed in both total intensity and polarized intensity in order to fully characterize disk phase function, polarizability curve, and scattering efficiency, ideally over the whole optical-NIR range (cometary and asteroid dust have red and blue "polarization colors", respectively). With such data in hand, it will be possible to perform direct comparisons to Solar System dust. Currently, the polarizability curve of debris disks is limited to small ranges of scattering angles (e.g, Tamura et al. 2006), with only the HR 4796A ring being fully characterized (Perrin et al. 2015). The ability to detect the back scattering peak in highly inclined disks (at scattering angles $\gtrsim160^\circ$), which requires high contrast and sensitivity at very small inner working angles, is critical to achieve this goal. 


\subsection*{Directly Testing the interaction between planets and debris disks}

Planets gravitationally sculpt debris disks and create large-scale structures, from rings (e.g., Kalas et al. 2005), warps (e.g., Heap et al. 2000), and other disturbed morphologies (Lee \& Chiang 2016) to distinct patterns like mean-motion resonant (MMR) structures (e.g.,Wyatt 2003).  These structures may help us constrain the planet's mass and orbit, the migration rate/grain size of dust, and indicate the presence of otherwise unobservable planets. 

\begin{wrapfigure}{l}{0.53\textwidth}
     \centering
    \includegraphics[width=0.52\textwidth]{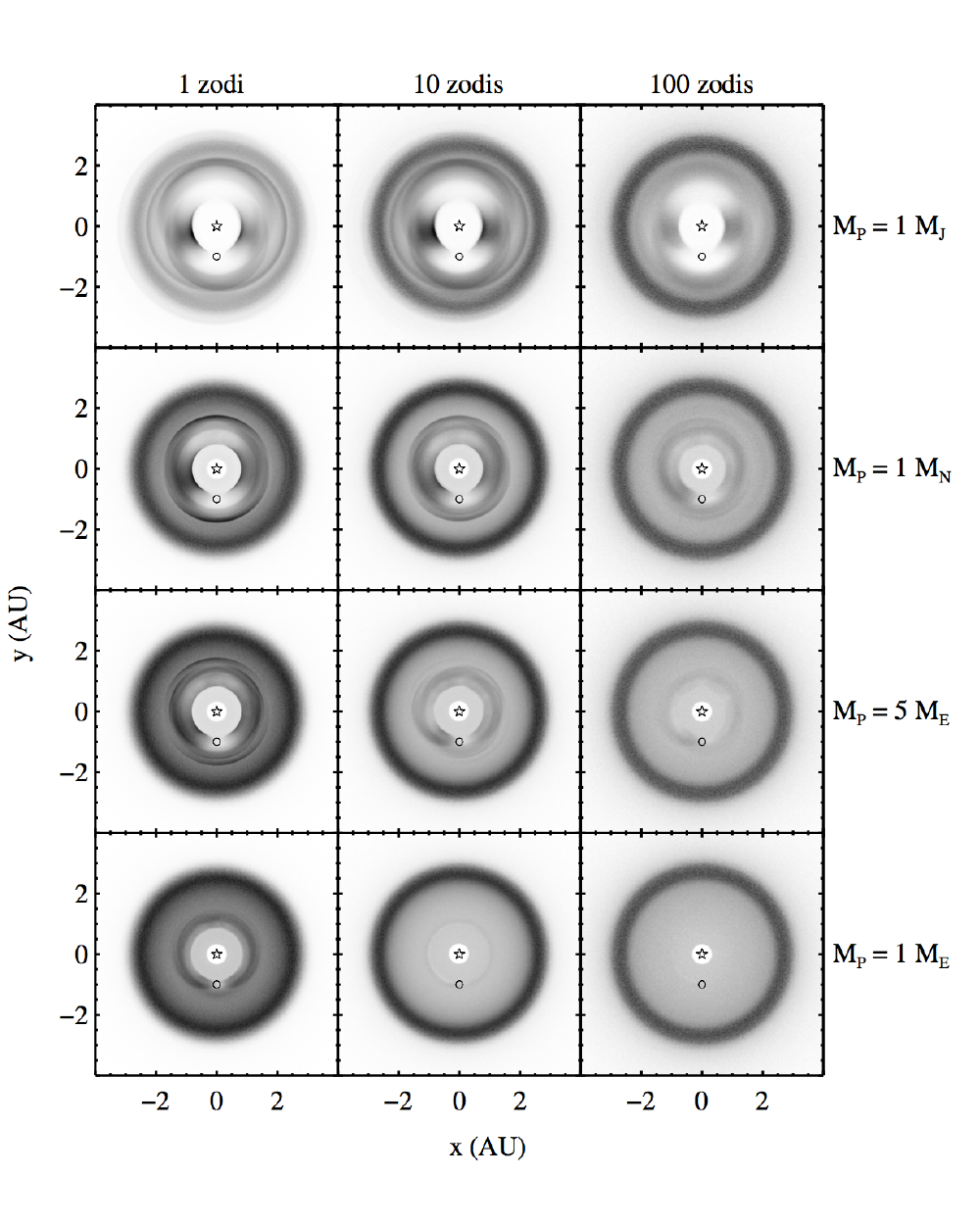}
    \caption{This figure, reprinted from Stark et al. (2011) demonstrates the impact that collisions have on mean motion resonance structures caused by companion planets. As there is more dust in a disk, structures are washed out by collisions.}
    \label{fig:f4}
\end{wrapfigure}

The use of disk structures as a signpost for planets has had limited results, primarily for two reasons.  Firstly, collision dominated disks tend to erase the more unique, easily-identifiable structures like MMR structures (Stark \& Kuchner 2009), leaving us with more ambiguous structures like gaps and warps that can be created by a variety of processes (e.g., Klahr \& Lin 2005, Lyra \& Kuchner 2013).  Second, because we are limited to disk structures at large stellar separations, where planets are faint in reflected light, we have only a single system for which planet-induced disk structure and the planet itself have been imaged: $\beta$~Pictoris (Dent et al. 2014), which is viewed edge-on and thus difficult to interpret.  


We recommend directly imaging the cold debris disks of known planet hosting stars with {\em Herschel}-detected disks of low luminosity which are likely to be dominated by drag processes and more amenable to showing MMR features. HR~8799, $\epsilon$~Eridani, and 51~Eri are also prime targets. New radial velocity surveys, GAIA astrometry, or other methods may open up a larger sample of disks to study disk-planet interactions within 100~pc.

\subsection*{Observational Requirements}

Our recommendations require a contrast ratio in the visible or NIR for total and polarized intensity light of $\sim$10$^{-10}$ for stars that have apparent V $<$10. This contrast must extend from 0.1$\arcsec$-30$\arcsec$\ to probe the innermost edges of cold debris disks ($\sim$5-10~AU at 100~pc) as well as to image disks of our nearest stellar neighbors ($\sim$100~AU at 3~pc). For large and nearby disks, spatial resolution can be $\sim0.5\arcsec$\ but more compact disks and MMR features require spatial resolutions of $\sim$50~mas to resolve features at our proposed inner working angle. Some of these requirements will be met by future ELT instrumentation, the WFIRST/CGI, WFIRST+starshade as well as the HabEx and LUVOIR concepts. Most future facilities have restricted outer working angles, and do not connect to the sensitivity of HST instruments at larger separations (i.e., Fomalhaut's disk). This gap can be addressed by space missions with small apertures and correspondingly large dark holes (e.g. Bryden et al 2011, Guyon et al 2012) or the use of a starshade and a large detector field-of-view. The feasibility of suborbital missions has been advanced by recent suborbital coronagraph and wavefront sensing missions such as PICTURE and HiCIBaS (Douglas et al 2018, Mendilllo et al 2017, Côté et al 2018).

\pagebreak
\noindent\textbf{References}
{\color{white}.}\\
Bailey et al. 2018, Proc. SPIE, 10698\\
Beuzit et al. 2019, Subm. to A\&A (arXiv:1902.04080)\\
Boccaletti et al. 2015, Nature, 526, 230 \\
Boley et al. 2012, ApJL, 750, L21\\  
Bryden et al. 2011, Proc. SPIE, 81511 \\
Côté et al. 2018, Proc. SPIE, 1070248 \\ 
Chen et al. 2014, ApJS, 211, 25;\\
Choquet et al., 2018, ApJ, 854, 53;\\
Debes et al. 2009, ApJ, 702, 318 \\
Debes et al. 2009, ApJ, 673, 191 \\
Dent et al. 2014, Science, 343, 1490 \\
Douglas et al. 2018, JATIS, 4, 019003 \\
Eiroa et al. 2013, A\&A, 555, A11\\
Engler et al. 2018, A\&A, 618, A151\\
Exoplanet Science Strategy report 2018\\
Guyon et al. 2012, Proc. SPIE, 84421S \\
Faramaz et al. in prep., 2019\\
Frattin et al. 2019, MNRAS, 484, 2198 \\
Heap et al. 2000, ApJ, 539, 435\\
Hughes et al. 2018, ARA\&A, 56, 541 \\
Jones \& Gehrz 2000, Icarus, 143, 338 \\
Kalas et al. 2005, Nature, 435, 1067-1070\\
Klahr \& Lin 2005, ApJ, 632, 1113 \\
Kolokolova et al. 2004, Comets II, 577 \\
Kolokolova et al. 2018, JQSRT, 204, 138 \\
Kuchner \& Stark 2010, AJ, 140, 1007 \\
Levasseur-Rigourd et al. 2007, P\&SS, 55, 1010 \\
Lisse et al. 2012, ApJ, 747, 93\\
Lisse et al. 2017, AJ, 154, 182\\
Lyra \& Kuchner 2013, Nature, 499, 184 \\
Marino et al. 2018, MNRAS, 479, 5423-5439  \\
Marino et al. 2019, MNRAS, 484, 1257-1269  \\
Mendillo, et al. Proc SPIE, 1040010 \\
Marshall et al. 2018, ApJ, 869, 10\\
Montesinos et al. 2016, A\&A, 593, A51\\
Perrin et al. 2015, ApJ, 799, 182 \\
Plavchan et al. 2019, Astro2020 Science White Paper\\
Ricci et al. 2015, ApJ, 798, 124\\
Sibthorpe et al. 2018, MNRAS, 475, 3046\\
Schmid et al. 2018, A\&A, 619, A9\\ 
Stark et al. 2011, AJ, 142, 123 \\
Tamura et al. 2006, ApJ, 641\\
Wyatt et al. 2003, ApJ, 598, 1321 \\
Zubko et al. 2016, P\&SS, 123, 63\\
\end{document}